\documentclass[useAMS,usenatbib]{mnras}
\usepackage{epsfig}
\usepackage{amssymb}
\usepackage{amsmath}
\usepackage{hyperref}
\usepackage{xcolor}
\usepackage{subcaption}
\usepackage{graphicx}
\usepackage{wrapfig}
\usepackage{dblfloatfix}

\title{Cosmic-ray transport and gamma-ray emission in M31}
\author[Do et al.]{Audrey Do$^{1}$, Matthew Duong$^{1}$, 
Alex McDaniel$^{1,2,3}$\thanks{E-mail: \href{mailto:armcdan@clemson.edu}{armcdan@clemson.edu}},\newauthor  Collin O'Connor$^{1}$, Stefano Profumo,$^{1,2}$\thanks{E-mail: \href{mailto:profumo@ucsc.edu}{profumo@ucsc.edu}}, Justine Rafael$^{1}$, Connor Sweeney$^{1}$,\newauthor and    Washington Vera III$^{1}$\\
$^{1}$Department of Physics, University of California Santa Cruz, 1156 High St., Santa Cruz, CA 95064, USA\\
$^{2}$Santa Cruz Institute for Particle Physics, 1156 High St., Santa Cruz, CA 95064, USA\\
$^{3}$Department of Physics and Astronomy,  Kinard Lab of Physics, Clemson, SC 29634-0978, USA}

\begin{document}

\maketitle

\begin{abstract}
    We study the possibility that an extended cosmic-ray leptonic and/or hadronic halo is at the origin of the large-scale gamma-ray emission detected from the Andromeda Galaxy (M31). We consider a broad ensemble of non-homogeneous diffusion scenarios and of cosmic-ray injection sources. We find that  cosmic-ray electrons and protons could be, and very likely are, responsible for part, or all, of the gamma-ray emission from M31, including out to more than 100 kpc from the center of the galaxy. We also simulate possible emission from pulsars in M31, and consider the effect of regions of highly inefficient diffusion around cosmic-ray acceleration sites, as suggested by recent TeV halo observations with Cherenkov telescopes.
\end{abstract}


\section{Introduction}
As the nearest large spiral galaxy to the Milky Way, M31 has been the subject of intense observational scrutiny, including recent detections at gamma-ray energies: While early gamma-ray telescopes were able to only set upper limits on the gamma-ray signal, the Fermi Large Area Telescope (LAT) \citep{Atwood:2009ez} was the first instrument to obtain a significant positive detection \citep{M31first}. The LAT detection was found to be compatible both with a point source and with an extended source emission tracing an infrared map at 100$\mu$m intended to indicate star-forming regions, with the extended emission preferred non-significantly at the confidence level of 1.8$\sigma$ \citep{M31first}. Subsequent studies of LAT data including longer exposure have added to the evidence for gamma-ray emission in M31, including the tentative, and controversial, detection of potential ``bubble-like'' features analogous to the Milky Way Fermi bubbles \citep{Pshirkov:2016qhu}. In a recent study, the Fermi-LAT collaboration reported a $10\sigma$ detection of M31 with a strong detection of spatially extended emission out to $\sim 5$ kpc at the $4\sigma$ significance level \citep{ackermann}. 

The nature and origin of the emission from the central regions of M31 remain somewhat controversial: on the one hand, M31's observed gamma-ray luminosity does not significantly deviate from the expectation from the known scaling relationship between infrared and gamma-ray luminosity \citep{Ajello:2020zna, Storm:2012gn}; this, in turn, would hint at cosmic rays, accelerated in supernova explosions, as the physical counterpart to the observed emission. This possibility was quantitatively explored in \cite{McDaniel:2019niq}, which found, however, that the required input power from supernova explosion would imply, in the case of a leptonic or hadronic, or even of a mixed scenario, a supernova rate around two orders of magnitude larger than expected. \cite{McDaniel:2017ppt} and \cite{McDaniel:2018vam} explored, instead, a dark matter annihilation scenario, where gamma rays originate as a result of the pair-annihilation of dark matter particles. This possibility was recently also considered in \cite{M31HaloDM}. While in principle consistent with the so-called, controversial, ``Galactic Center Excess'' in the Milky Way \citep{fermiGCE}, and marginally in tension with the non-observation of gamma-ray emission from local dwarf galaxies by Fermi-LAT \citep{albert2017, fermi_dwarfs2015}, this is an intriguing possibility. Finally, unresolved emission from point sources such as millisecond pulsars or other compact objects, which has been considered in \citet{fragione} as well as in \citet{eckner}, remains an unavoidable component of the observed signal, albeit with uncertain relative importance.

Other recent observations of gamma-ray emission in M31 have searched for emission at large radii in the outer halo of the galaxy. As part of a detailed study of the gamma-ray emission in M31 using roughly 8 years of Fermi-LAT data in a $60^{\circ}$ region of interest centered at $(l,b) = (121.17^{\circ}, -21.57^{\circ})$,  \citet{M31Halo} reports evidence for an extended gamma-ray excess separate from the Milky Way foreground. This purported emission extends out to roughly 100-200 kpc above the plane of the galaxy, although the authors acknowledge that the emission from the ``far outer halo'' (at angles from M31's center $8.5^\circ<r<21^\circ$) is likely related to mis-modeling of the significant foreground emission from the Milky Way and thus less robust than the emission from the ``Spherical Halo'' region at angles between $0.4^\circ<r<8.5^\circ$ and than the  robustly-detected inner galaxy emission at $r<0.4^\circ$. \citet{M31Halo}, while not ruling it out, argues against an extended cosmic-ray halo \citep{Feldmann:2012rx, Pshirkov:2015oqu} based on the radial extent, spectral shape, and intensity of the observed large-radii signal. However, as we argue below, the radial extent and intensity depend critically on assumptions on cosmic-ray diffusion outside the Galactic plane and in the halo; and the spectral shape is strongly affected by foreground Galactic emission and from the intrinsic weakness and limited statistics of the signal.

In this study, we study cosmic-ray electron and proton transport in M31 under a variety of assumptions on the nature of diffusion within and beyond the traditional cylindrical ``diffusion box'', used in Milky Way cosmic-ray studies, around M31's galactic plane. Since we relax the simplifying assumption of homogeneity for the diffusion coefficient, we solve the transport equation via a stochastic approach in the standard way (namely turning the Fokker-Planck partial differential equation describing cosmic-ray transport into a stochastic differential equation solved by means of a Monte Carlo method). We consider both a sharp discontinuity and a gradual transition from the inside to the outside of the inner diffusion region; in addition, we also consider a model, that is becoming increasingly well-motivated by observations of TeV halos \citep{Abeysekara:2017old}, where diffusion within the sites of cosmic-ray acceleration is inefficient. Finally, we also consider a variety of possible injection sites for the cosmic rays.

Cosmic-ray electrons and protons both produce gamma rays as they propagate through the galaxy. However, while electrons radiate highly efficiently and lose energy quickly, protons' energy losses are significantly less efficient, with time scales much longer than those associated with propagation. At the gamma-ray energies of interest, and in the outer regions we are concerned with, cosmic-ray electron emission proceeds through inverse-Compton scattering, primarily by up-scattering photons in the cosmic microwave background (CMB). Cosmic-ray protons instead produce gamma rays as a result of inelastic collisions with the interstellar and circumgalactic medium, producing neutral pions eventually decaying to gamma-ray pairs.

In the study below we also re-assess the contribution of millisecond pulsars and of younger pulsars to the gamma-ray emission, making use of dedicated pulsar population synthesis modeling and of observationally-motivated predictions for gamma-ray emission from pulsars.

Our results indicate that it is quite plausible that {\em (i) most of the spherical halo gamma-ray emission originate from a cosmic-ray halo possibly extending out to M31's virial radius, and well beyond the galactic disk}; we find that this interpretation is {\em (ii) possible both within hadronic and leptonic cosmic-ray scenarios}, albeit in the latter case only a fraction of the spherical-halo gamma-ray emission can be explained; finally, we find that the {\em (iii) inner-galaxy emission is most likely a combination of pulsar gamma-ray emission and of hadronic and leptonic cosmic-ray-induced gamma rays}, somewhat attenuating the tension with the expected supernova rate found in \cite{McDaniel:2019niq}.

The remainder of this study is structured as follows: in the next section \ref{sec:diff} we outline our approach to solving the relevant transport equation, and give details about the diffusion and cosmic-ray models we consider; the following sections \ref{sec:results} and \ref{sec:psr_gamma} detail our results on cosmic-ray driven gamma-ray emission and on pulsar gamma-ray emission, respectively; finally, sec.~\ref{sec:conclusions} presents a final discussion of our results and our conclusions.

\section{Solution to the transport equation and diffusion models}\label{sec:diff}
The transport of cosmic rays on galactic scales, describing the particles' flux and energy spectrum $n(\vec x,\vec p,t)$, with $\vec x$ position and $\vec p$ momentum, is customarily described through diffusive processes in phase space via equations with a structure of the type \citep{Strong:1998pw}
\begin{multline}
    \frac{\partial n}{\partial t}+\vec u\cdot\nabla n=\nabla\cdot(\hat \kappa\nabla n)+\frac{1}{p^2}\frac{\partial}{\partial p}\left(p^2\kappa_{pp}\frac{\partial n}{\partial p}\right)\\
    +\frac{1}{3}(\nabla\cdot \vec u)\frac{\partial n}{\partial \ln p}+S(\vec x,p,t).
\end{multline}
In the equation above, $\vec u$ is the advection speed, $\hat\kappa$ the spatial diffusion tensor, $p=|\vec p|$, $\kappa_{pp}$ is the momentum-space diffusion coefficient which effectively describes re-acceleration, and $S(\vec x,p,t)$ describes the cosmic-ray sources.

In the context of cosmic-ray (CR) transport, the diffusion equation has been solved through a variety of methods. In \citet{Colafrancesco:2005ji,Colafrancesco:2006he}, a Green's function method, with image ``charges'' suitably accounting for the boundary conditions of the problem, was developed and employed to solve for the steady-state solution to a diffusion problem with spherical symmetry, for arbitrary injection spectra, but with spatially constant energy loss term ad diffusion coefficient \citep[see also][]{2020arXiv201111947V}. The method was generalized in \cite{McDaniel:2017ppt} for the calculation of the radio and inverse-Compton emission, with the possibility to also include a spatially-dependent magnetic field and target radiation field energy density.

An alternate approach is to solve for the differential equation on a lattice by discretizing the problem in a standard fashion. This method is employed by popular codes that solve for the diffusion problem in cylindrical coordinates such as GALPROP \citep{Vladimirov:2010aq} or DRAGON \citep{Tomasik:2008fq}. While this method can in principle be adapted to different geometries and to different assumptions on the spatial dependence of the various transport coefficients, the method is not easily adapted to complex diffusion setups such as the ones we are interested in here.

Finally, using the well-known connection between the Fokker-Planck equation and Stochastic differential equations, other codes model CR transport by means of stochastic processes, see for example the CRPropa code \citep{Merten:2017mgk, Batista:2019nbw}. Here, we utilize precisely this approach, as it is the most flexible to study largely inhomogeneous diffusion setups and complex CR injection morphologies. We refer the reader to classic literature on the equivalence between Fokker-Planck partial differential equations and stochastic differential equations, see e.g. \cite{23677}.

\begin{figure*}
\centering
    \includegraphics[width=0.44\linewidth]{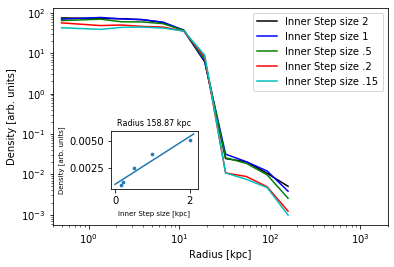}
    \includegraphics[width=0.48\linewidth]{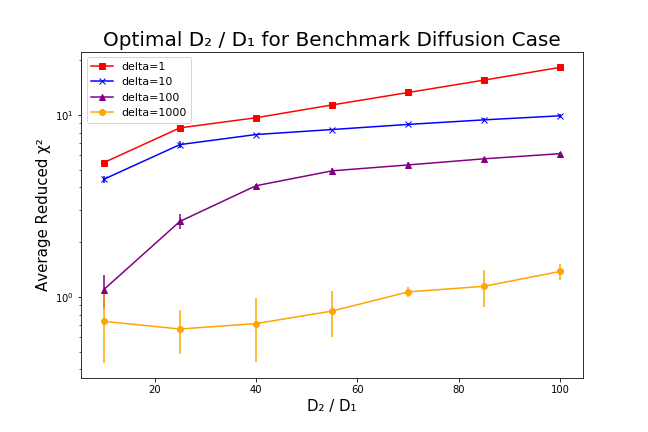}
    \caption{Left: Illustration of an extrapolation procedure used in some of our simulations: The inner step size refers to the step size within the diffusion region. This image represents the CRP1 diffusion setup. Right: Averaged $\chi^2$ for a fit to the observed gamma-ray emission morphology from cosmic-ray protons with an ensemble of simulation at various values of the width of the transition region $\delta$ and of the ratio $D_2/D_1$. The optimal choice corresponds to $D_2/D_1=25$ and $\delta=1000$ kpc.}
    \label{fig:d2d1}
\end{figure*}

We model diffusion as a stochastic process in space (we model energy losses separately, and neglect reacceleration). We assume diffusion to be isotropic, thus each pseudo-particle's step is taken to occur in a random direction in space. The step size is taken to correspond to the mean free path $\lambda$, which for a diffusive process in 3 dimensions is $\lambda\simeq 3D/v$, with $D$ the (energy-dependent) diffusion coefficient at the particle's location, and $v$ the pseudo-particle velocity; for instance, for typical values of $D\sim3\times 10^{28}\ {\rm cm}^2/s$ and $v\simeq c$, we get $\lambda\simeq 1$ pc. Since the diffusion coefficient is taken to be a function of energy, so is the corresponding mean free path. To reduce the computational complexity of our simulations, we occasionally needed to resort to extrapolations of the results of simulations with larger step sizes. The extrapolation procedure involves running several simulations with the same parameters, except with different step sizes. At each step size, there is a density of particles for a given radius. The densities are extrapolated as a function of step size and a best fit line is produced from the results. We then extrapolate the density to a step size of .002 using the best fit line. A visual of this procedure is shown in fig.~\ref{fig:d2d1}, left.

To model the CR spatial distribution in steady state (the case of interest here, since we assume all injection sources to be in steady state), we run our simulations with a limited number of pseudo-particles and assess the CR residence time-scale $\tau$ for pseudo-particle loss outside the region of interest (which is typically taken to be 200 kpc, the same region of interest used in \cite{M31Halo}, and around 2/3 of M31's virial radius) by fitting, after an initial transient, for the exponential decay behavior of the number $N(t)$ of pseudo-particles still within the diffusion region versus the total initial number of particles $N(t=0)=N_0$,
\begin{equation}
    N(t)\simeq N_0\exp(-t/\tau);
    \label{eq:particle number}
\end{equation}
In practice, we fit the exponential form to the interval  $0.1<N(t)/N_0<0.9$ to prevent both fitting for the initial transient corresponding to the drift to the boundary of the diffusion region, and for the noisy tail of the distribution at the end of the simulation. Once the residence time-scale $\tau$ is found, we run extensive simulations with a large number of pseudo-particles (on the order of $10^6$ to $10^8$) for a time $t=\tau$.

We validated the procedure outlined above and the code by comparing our results with a simplified version of the diffusion equation's Green function, which solves
\begin{equation}
    \frac{\partial n}{\partial t}=\nabla\cdot (\hat\kappa\nabla n)+\delta(\vec r)\delta(t)
\end{equation}
and which reads
\begin{equation}
    n(r_i,t)=\frac{\exp\left(-\frac{r_i^2}{4\kappa_{ii}t}\right)}{\left(4\pi \kappa_{ii}t\right)^{d/2}},
\end{equation}
with $d$ the dimension and $\kappa_{ii}$ the diffusion coefficient in the direction $r_i$. We cross-checked our code for $d=1,\dots,3$ and for isotropic and non-isotropic diffusion tensor. We completely developed the code we employed in-house.

We neglect energy losses for the case of cosmic-ray protons, while in the case of cosmic-ray electrons we assume the standard quadratic dependence on energy for energy losses for high-energy electrons,
\begin{equation}
    \frac{dE}{dt}\simeq -b_0 E^2,
\end{equation}
with $b_0\sim 10^{-16}\ {\rm GeV}^{-1}{\rm sec}^{-1}$ \citep{Colafrancesco:2005ji}. One can integrate the equation above by separation of variables between an initial and a final time/energy to get (for initial time $t=0$)
\begin{equation}
    -\frac{1}{E_f}+\frac{1}{E_i}=-b_0t=-b_0x/c,
\end{equation}
where in the last equation we assumed that the propagation step has length $x=ct$ because the electrons are ultra-relativistic. Solving for $E_f$, one finds:
\begin{equation}\label{eq:energyloss}
    E_f=\frac{E_i}{1+\left(\frac{b_0}{c}\right)x\ E_i}.
\end{equation}
Expressing $x$ in kpc and $E_i$ in GeV, the multiplicative constant gives
\begin{equation}
E_f=\frac{E_i}{1+10^{-5}\frac{x}{\rm  kpc}\ \frac{E_i}{\rm  GeV}}.
\end{equation}
Thus, in the case of the electrons, we run simulations that track not only position, but also energy. We assume an injection spectrum inspired by Fermi second-order acceleration, $dN/dE\sim 1/E^2$, and 
spawn electrons with energy in proportion to that spectrum; we then track both where the final position of the electrons is and which energy they have, with the proviso that, for every step where the electron has moved by a distance $x$,  the energy is reduced according to the equation (\ref{eq:energyloss}) above.

Notice that while we diffuse electrons in the same way as protons, i.e. simulating diffusion via a stochastic processed as described above, the relevant time scale for the simulation in this case is the {\em energy-loss} time scale rather than the diffusion of the particles outside the diffusive region of interest. The energy-loss time scale is defined by the number of steps it takes particles to fall below an energy such that emission of gamma rays in the energy range to which the Fermi LAT is sensitive to is no longer viable. Since we are interested here primarily in emission far away from the stellar population in M31, the main photon field we are concerned with is the cosmic microwave background (CMB) (the inverse-Compton emission scales with the energy density in a given radiation field, and in the spherical halo this is dominated by the CMB). The average energy of a CMB photon is $E_\gamma\simeq 6.4\times 10^{-4}$ eV; the typical energy of the up-scattered photon is $E_f\sim \gamma_e^2 E_\gamma$, where $\gamma_e$ is the Lorentz factor of the incoming electron. Requiring $E_f\gtrsim 0.1$ GeV, which is at the lowest energy detectable by the LAT, we get that $\gamma_e=E_e/m_e\gtrsim \sqrt{E_f/E_\gamma} \simeq 4\times 10^5$, thus $E^{\rm min}_e\simeq 200$ GeV. Notice that the energy dependence of the diffusion coefficient is quite critical in the case of high-energy electrons, $D(E)=D_0(E/{\rm GeV})^\delta$. We assume $\delta\simeq 0.3$.

We can estimate the average path length $\lambda_e$ the electrons take to lose their initial energy $E_i$ from Eq.~(\ref{eq:energyloss}) above; being a diffusive process, we first calculate the time $T_{i\to f}$ for the electrons to lose energy from $E_i$ to $E_f$
\begin{equation}
    T_{i\to f}=\frac{1}{b_0}\left(\frac{1}{E_f}-\frac{1}{E_i}\right)\simeq 10^{16}\frac{\rm GeV}{E_f}\ {\rm sec};
\end{equation}
The typical distance traveled by an electron from its injection point is then given by
\begin{equation}
    \lambda_e\simeq\sqrt{D_0(\bar E/{\rm GeV})^\delta T_{i\to f}}\sim 5.8\ {\rm kpc}\ \left(\frac{E_f}{\rm GeV}\right)^{-0.35}\approx 0.9\ {\rm kpc},
\end{equation}
where in the last equation we assumed $E_f\simeq 200$ GeV. Thus, electrons contributing IC emission off of CMB are not expected to diffuse further than approximately 1 kpc from the source location. Our simulations are found to be consistent with the simple estimate above.

While in the case of electrons up-scattering CMB photons there is a one-to-one correspondence between the final electron locations and the gamma-ray emission along a given line of sight, in the case of protons the structure of the interstellar and ciscumstellar target gas density is crucial, as the gamma-ray emissivity is proportional to the line-of-sight integral of the product of the cosmic-ray proton density and the target gas density. Given the lack of detailed information about the gas density along the line of sight especially between the MW and M31 centers, we choose to adopt the results of the simulations in \cite{2014MNRAS.441.2593N}. Specifically, we utilize their results on the gas density along the line of sight between M31 and the MW (thick black line in fig.~16), up to a distance of approximately 30 kpc from M31; at that point, we match the gas density in their simulation results for the gaseous halo of M31, fig.~5, left panel; finally, we use the results of the simulations in \cite{2013ApJ...763...21F}, their fig.~1, summing upon all components, in the innermost 1.5 kpc. 

In producing the gamma-ray morphology plots, we simply count the number of pseudo-particle corresponding to cosmic-ray electrons, given the homogeneity of the background radiation field. In the case of cosmic-ray protons we instead weigh each pseudo-particle with the corresponding target gas density at its location, and then integrate (i.e. sum) along the given line of sight.

In order to compare our results with observations, we digitized the map corresponding to the tentative signal (i.e. residual) intensity map in \cite{M31Halo}, fig. 34, top-left,  using the numpy, matplotlib, and OpenCv libraries in python to digitally input the image and the associated color-bar, which was then converted from RGB values back into physical values using the nearest color index of the color-bar. This was then scaled down to the source's original 20x20 pixel resolution. We note that the source image contained overlays for spatial reference, which resulted in artifacts upon digitizing and re-scaling; in order to resolve this, pixels were manually corrected by sampling the nearest unaffected pixels from the full-scale digitized image.

\begin{figure*}
    \centering
    \mbox{\includegraphics[width=0.3\linewidth]{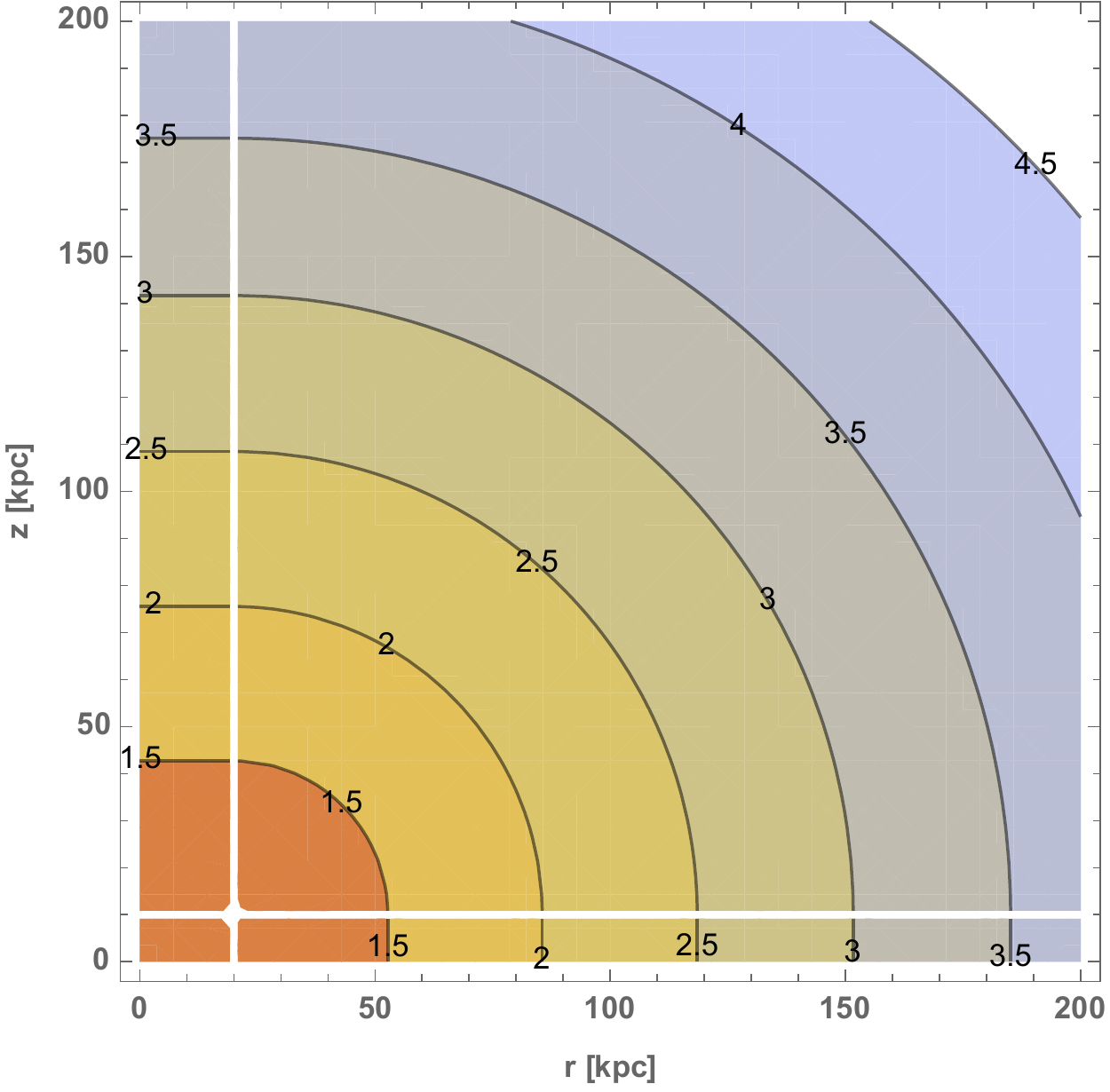}\qquad\qquad \includegraphics[width=0.4\linewidth]{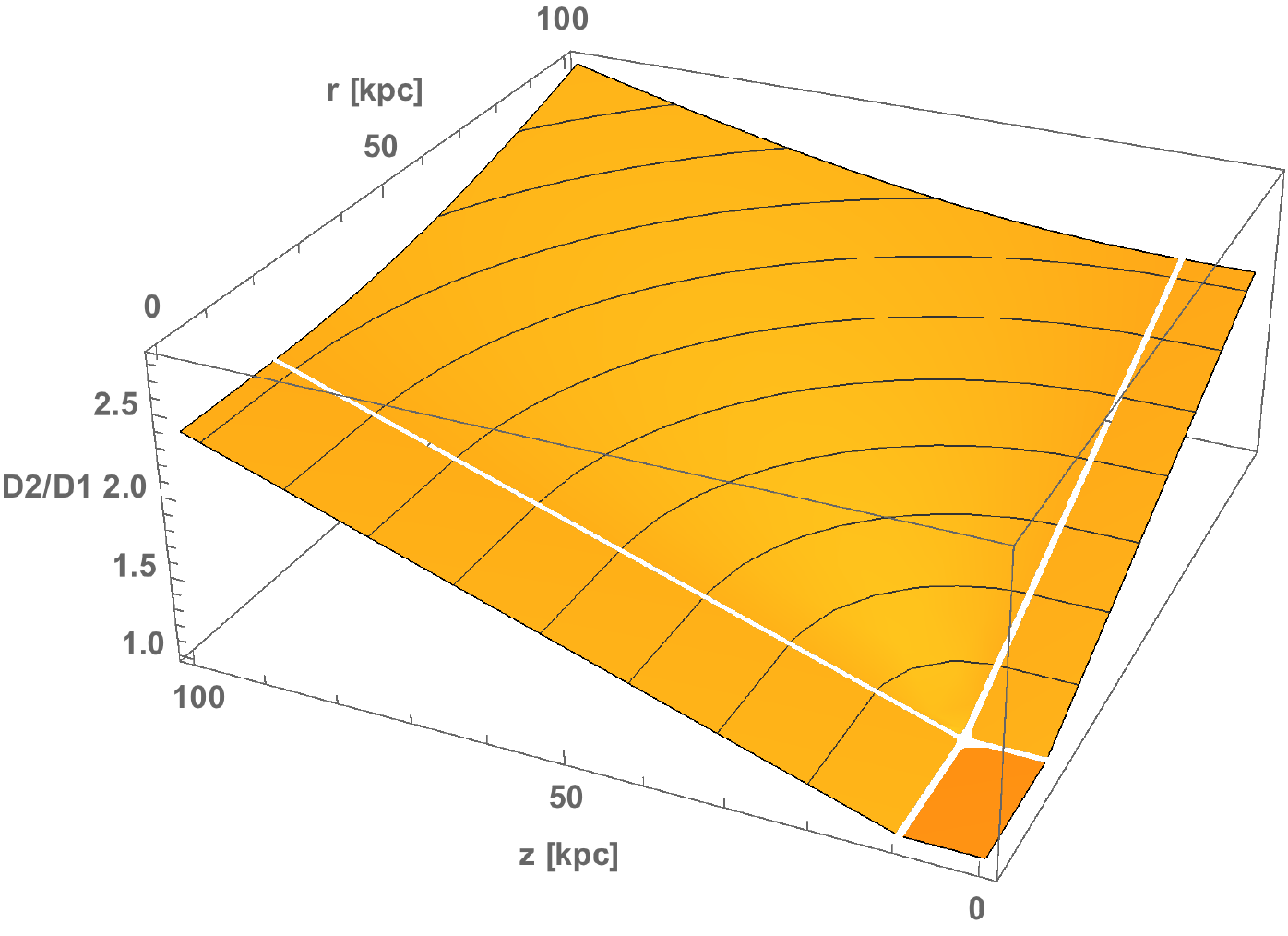}}
    \caption{Spatially varying diffusion coefficient, with the functional form in Eq.~(\ref{eq:diff}) and $D_2/D_1=25$ and $\delta=1000$ kpc. The left panel shows curves of constant $D_2/D_1$ on the $(r,z)$ plane of cylindrical coordinates across the entire region of interest, while the right panel shows a three dimensional plot of the same quantity, on the same plane. The inner diffusion region, with constant $D=D_1$, is bordered by white lines.}
    \label{fig:diffcoeff}
\end{figure*}

\section{Results}\label{sec:results}
In this analysis we test the impact of both the source morphology and the diffusion setup on the gamma-ray emission from (i) inelastic collisions of cosmic-ray protons with the ISM and (ii) inverse Compton emission of high-energy cosmic-ray electrons off of CMB photons, using the simulation techniques described in the previous section.

As far as the source morphology is concerned, we entertain three scenarios:
\begin{enumerate}
    \item Cosmic rays are produced near the central region of M31; this scenario assumes that the main acceleration mechanism for cosmic rays in M31 is physics associated with the innermost region of the galaxy, such as for instance accretion around, and jets emanating from, the central supermassive black hole of M31;
    \item Cosmic rays are produced in star-forming regions; this possibility physically relies on the notion that the main cosmic-ray acceleration sites are likely supernova shocks, whose locations trace star-forming regions. Observationally and theoretically, this possibility was explored in \cite{Carlson:2015daa, Carlson:2015ona} and in \cite{Carlson:2016iis}, which found that a significant fraction of cosmic rays in the Milky Way are likely injected from star-forming regions. We use as a tracer of star-forming regions in M31 the IR emission map from \cite{Gordon_2006}.
    \item Finally, we use the {\tt PrsPopPy} code (\cite{psrpoppy}) to produce a synthetic population of pulsars (see sec.~\ref{sec:psr_gamma} for details on the population synthesis procedure we adopt) as a proxy for a scenario where cosmic-ray electrons and positrons are produced in the magnetosphere of rotating neutron stars \citep[see e.g.][]{Grasso:2009ma, Profumo:2008ms}.
\end{enumerate}

We also entertain a variety of diffusion models, taking advantage of the flexibility provided by the stochastic solution to the diffusion equation. In particular, we assume:
\begin{enumerate}
    \item[(a)] A traditional ``leaky box'' diffusion scenario (hereafter referred to as our ``benchmark'' model) inspired by similar setups for the Milky Way that successfully reproduce the measured abundance of cosmic-ray species \citep[see e.g.][]{Vladimirov:2010aq, Tomasik:2008fq}. Here, we assume that cosmic rays diffuse primarily inside a cylindrical diffusion region of radius 20 kpc and half-height 10 kpc (we have also considered variations of these parameters, with marginal impact on our results described below), effectively free-streaming outside the diffusion region; we model this latter effect by a sudden, step-like jump by a factor of 100 in the diffusion coefficient outside the cylindrical box;
    \item[(b)] A ``constant'' diffusion scenario, where cosmic rays diffuse in an isotropic and homogeneous medium with a constant diffusion coefficient. Albeit physically unrealistic, this scenario aims at assuming that the circumgalactic medium in the Local Group continues to support cosmic-ray diffusion well outside M31 and the Milky Way and out to much larger radii than the galaxies' size;
    \item[(c)] A ``gradual'' spatially-dependent diffusion coefficient defined so that the diffusion coefficient inside a cylindrical box of height $z_t$ and radius $r_t$ is $D_1$ and, after a transition region of size $\delta$, it asymptotes to an outer value $D_2$. The distance of a point $(r,z)$ in cylindrical coordinates from the diffusion box is
    \begin{equation}
        {\rm dist}(r,z)=\sqrt{(r-{\rm min}[r,r_t])^2+(z-{\rm min}[z,z_t])^2};
    \end{equation}
    The diffusion coefficient at a point of cylindrical coordinates $(r,z)$ is then calculated as
    \begin{equation}\label{eq:diff}
        D(r,z)=D_1+\left(D_2-D_1\right)\frac{{\rm ArcTan}\Big[{\rm dist}(r,z)/\delta\Big]}{\pi/2}.
    \end{equation}
    We searched for the values of $D_2/D_1$ and $\delta$ producing the gamma-ray emission morphology most closely resembling (based quantitatively on a pixel-by-pixel $\chi^2$ procedure) the diffuse gamma-ray emission measured in the inner halo and spherical halo of M31 \citep{M31Halo}. For this calculation, we utilized the gamma-ray emission from cosmic-ray protons, with the procedure explained above. The $\chi^2$ was computed by comparing the predicted and measured emission pixel by pixel, after normalizing both maps to the same average emission. We show in fig.~\ref{fig:d2d1}, right, the results for the $\chi^2$ for different values of $\delta$ and $D_2/D_1$. 
    For every combination of $D_2/D_1$ and $\delta$ we ran a set of 10 independent simulations, with the inferred standard deviation shown in the figure. Our results indicate a preference for {\em large} values of $\delta$, implying, in turn, a preference for a mild ``gradient'' in transitioning to the larger outer value of the diffusion coefficient. Similarly, we observe a preference for smaller ratios of the outer to inner diffusion coefficient. However, for large $\delta\sim1$ Mpc, we find that ratios $5\lesssim D_2/D_1\lesssim 50$ give equally good fits to the observed morphology. While there is no strong statistical preference, we adopted as our benchmark choice the lowest $\chi^2$ central value which corresponded to $D_2/D_1=25$. We show in     fig.~\ref{fig:diffcoeff} with an iso-level contour plot in the left and with a three-dimensional rendering of a 100$\times$100 kpc region on the right the resulting diffusion coefficient (normalized to the value inside the inner diffusion box) in cylindrical coordinates $(r,z)$, with parameters corresponding to the optimal choices $D_2/D_1=25$ and $\delta=1000$ kpc.

\begin{figure*}
    \centering
    \begin{subfigure}[b]{1\textwidth}
    \includegraphics[width=\linewidth]{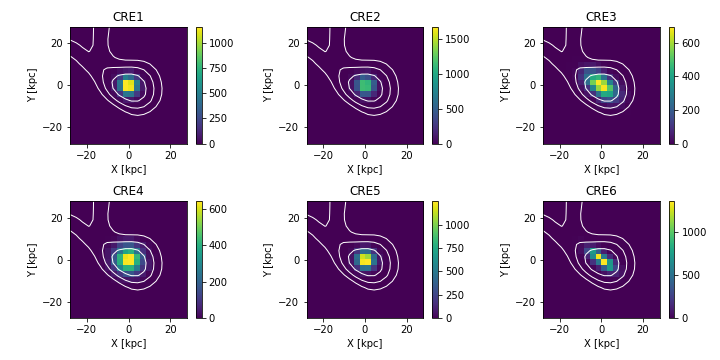}
    \end{subfigure}
    \caption{Morphology of each of the six CRE cases. For each case, a contour plot of the observed gamma-ray emission from the Fermi-LAT observations \citep{M31Halo} is mapped over the CRE configurations for comparison.}
    \label{fig:cre_morphology}
\end{figure*}
    
\begin{figure}
    \centering
    \includegraphics[width=\linewidth]{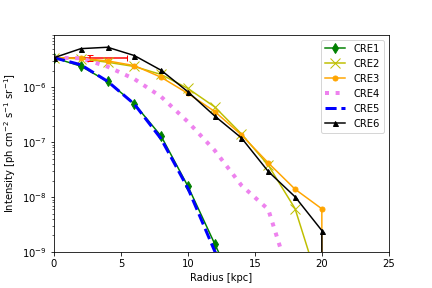}
    \caption{Radial intensity profile for cosmic-ray electron simulations, for the six diffusion and injection source profile combinations discussed in the text.}
    \label{fig:cre_radial_intensity}
\end{figure}

    \item[(d)] As a second example of a spatially-varying diffusion coefficient, we utilize a model (which we dub ``Swiss cheese'' diffusion coefficient) where inside the diffusion region there exist spherical sub-regions of inefficient cosmic-ray transport associated with the turbulent medium inside pulsar wind nebulae (PWNe). This scenario reflects the recent findings of \cite{Abeysekara:2017old} that high-energy cosmic-ray electrons diffuse much less efficiently (around a factor 100 smaller effective diffusion coefficient) inside PWNe than outside. Following  \cite{Profumo:2018fmz}, we use the model of \cite{Abdalla:2017vci} to relate the pulsar age to the radial size of the corresponding PWN, and we assume a sudden transition to a diffusion coefficient $D_0/100$ inside the PWN; outside the cylindrical box, we assume, as for the benchmark model, a large diffusion coefficient $100\times D_0$.
\end{enumerate}

We employ slightly different sets of source distribution and diffusion models for cosmic-ray electrons (CRE) and protons (CRP), based on different expected injection sources (protons are not thought to be produced by pulsars' magnetospheres). We describe below our choices and results.

\subsection{Cosmic-Ray Electrons}
Here we present results for gamma rays from inverse Compton (IC) up-scattering of CMB photons by high-energy cosmic ray electrons. We consider six different cases:
\begin{enumerate}
    \item[(CRE1):] Benchmark diffusion scenario (a), with CRE injected at the very center of M31, i.e. scenario (i)
    \item[(CRE2):] Benchmark diffusion scenario (a), with CRE injected in star-forming regions, i.e. scenario (ii)
    \item[(CRE3):] Benchmark diffusion scenario (a), with CRE injected at the location of mature synthetic pulsar locations, i.e. scenario (iii)
    \item[(CRE4):] Constant diffusion scenario (b), with CRE injected at the very center of M31, i.e. scenario (i)
    \item[(CRE5):] Gradual diffusion scenario (c), with $D_2/D_1=25$ and $\delta=1,000$ kpc, with CRE injected at the very center of M31, i.e. scenario (i)
    \item[(CRE6):] ``Swiss cheese" diffusion scenario (d), with CRE injected at the location of mature synthetic pulsar locations, i.e. scenario (iii)
\end{enumerate}

\begin{figure*}
\centering
    \begin{subfigure}[b]{1\textwidth}
    \includegraphics[width=\linewidth]{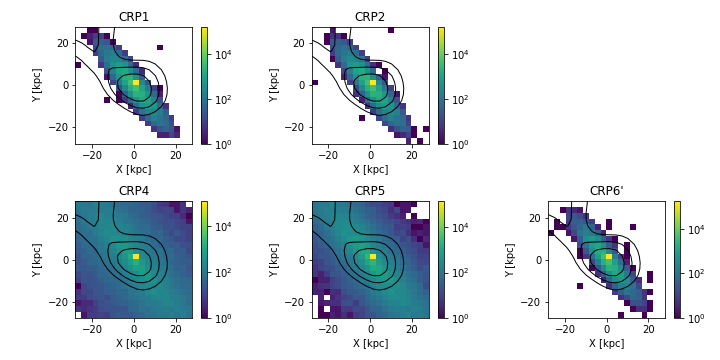}
    \end{subfigure}
\caption{Morphology of each of the five CRP cases. For each case, a contour plot of the observed gamma-ray emission from Fermi-LAT \protect\citet{M31Halo} is mapped over the CRP configurations for comparison.}
\label{fig:crp_morphology}
\end{figure*}

\begin{figure}
    \centering
    \includegraphics[width=0.85\linewidth]{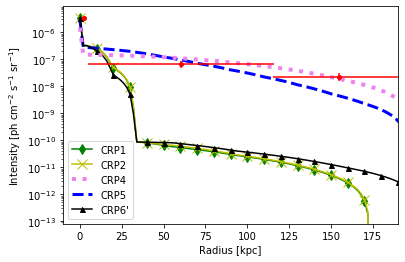}
    \caption{Radial intensity profile for cosmic-ray proton simulations, for the five diffusion and injection source profile combinations discussed in the text.}
    \label{fig:crp_radial_intensity}
\end{figure}

Fig.~\ref{fig:cre_morphology} shows the emission morphology from IC of CMB photons by CRE. We notice that virtually in all cases the emission is mostly circumscribed to the inner regions of M31, albeit with different morphology for the different assumptions on diffusion and source location. The radially-averaged intensity profiles of the six cases is shown in fig.~\ref{fig:cre_radial_intensity}. First, we note that CRE1 and CRE5 are very similar, indicating that the transport conditions beyond the inner regions play a relatively mild role. In these cases, there is also very marginal emission in the spherical halo region. CRE2 and CRE3 also look remarkably similar, as is somewhat expected since in both cases the CR injection sites trace star formation, in CRE2 via the IR emission map we utilize as a proxy for the star formation rate, and in CRE3 via the synthetic pulsar population model we constructed (described in detail in the following section). CRE2 and CRE3 exhibit a more extended morphology compared to CRE1 and CRE5, and are found to contribute somewhat to the emission in the spherical halo region, which is around $5\times 10^{-8}$ ph cm$^{-2}$ s$^{-1}$ sr$^{-1}$, out to around 15 kpc. Finally, CRE6 is the model that has the most pronounced emission in the innermost few kpc, driven by CR electrons being ``trapped'' in bubbles of inefficient transport associated with PWNe.

\subsection{Cosmic-Ray Protons}

    \begin{figure*}
    \centering
    \includegraphics[width=0.4\linewidth]{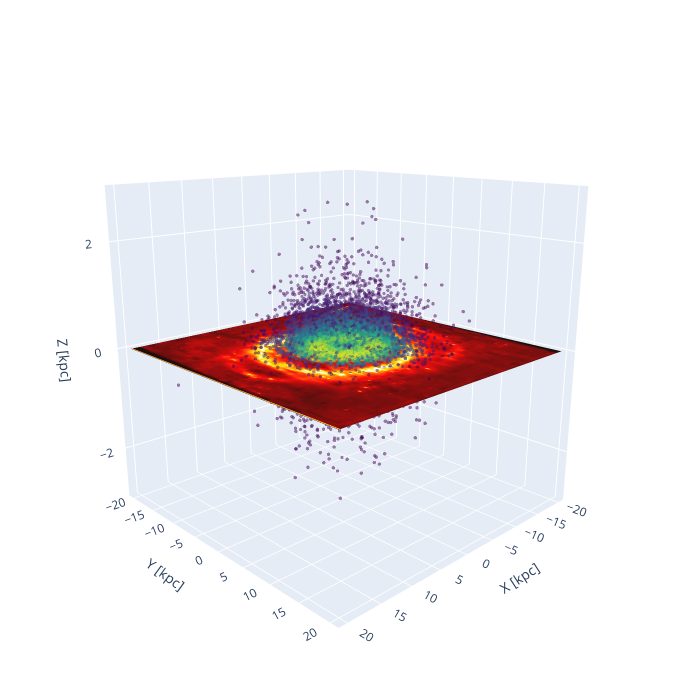}\qquad \includegraphics[width=0.4\linewidth]{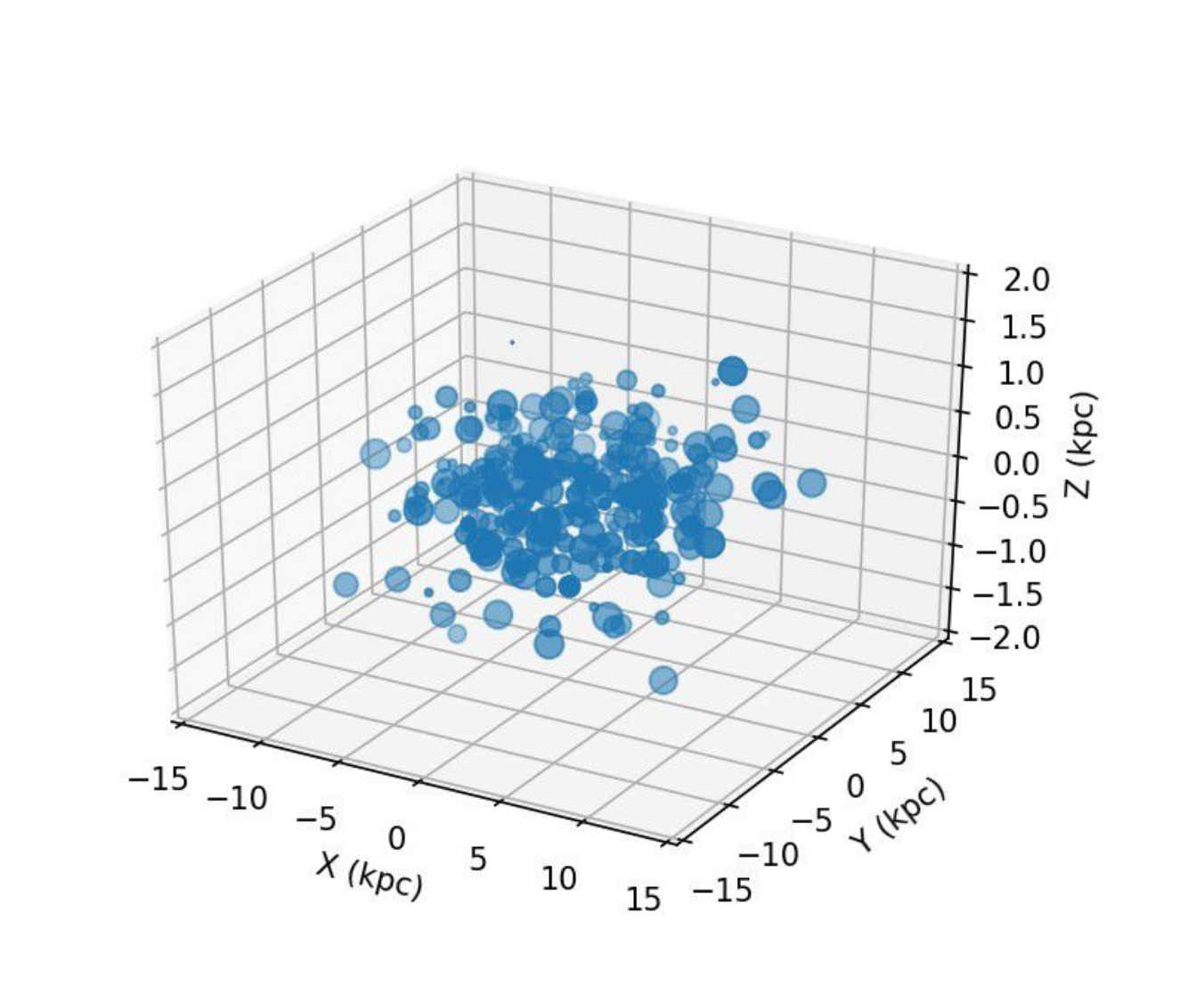}
    \caption{Left: Model pulsar population locations in relation to the galactic plane of M31. The colors of the pulsars are a generated by a standard 3-D kernel density estimation. The plane image is the de-projected 24 $\mu$m image from \protect\citet{Gordon_2006}. Right: example sample of 303 PWNe (enlarged by a factor 20).}
    \label{fig:psr_locations}
\end{figure*}

In the case of protons, which are not produced in pulsars' magnetospheres, we consider a different set of cases (although, to ease the comparison with the CRE case, we follow a similar numbering convention), specifically: 
\begin{enumerate}
    \item[(CRP1):] Benchmark diffusion scenario (a), with CRP injected at the very center of M31, i.e. scenario (i)
    \item[(CRP2):] Benchmark diffusion scenario (a), with CRP injected in star-forming regions, i.e. scenario (ii)
    \item[(CRP4):] Constant diffusion scenario (b), with CRP injected at the very center of M31, i.e. scenario (i)
    \item[(CRP5):] Inhomogeneous ``gradual'' diffusion scenario (c), with $D_2/D_1=25$ and $\delta=1,000$ kpc, with CRP injected at the very center of M31, i.e. scenario (i)
    \item[(CRP6$^\prime$):] ``Swiss cheese" diffusion scenario (d), but with CRP injected in star-forming regions, i.e. scenario (ii)
\end{enumerate}
As above, we show in two separate figures the results for the morphology of the innermost 25$\times$25 kpc region in fig.~\ref{fig:crp_morphology} and the radial intensity profile in fig.~\ref{fig:crp_radial_intensity}. Our results indicate that CRP1, CRP2, and CRP6$^\prime$ all exhibit a relatively similar morphology, likely due to the fact that in those cases the emission tracks quite closely the residence time, in turn related to the diffusion coefficient. Since protons diffuse for much longer times than electrons, the source injection site is less critical, and information thereof is asymptotically lost.

Larger values of the diffusion coefficient in the spherical halo and outer halo regions, as in CRP4 and CRP5, yields, as expected, a much brighter emission at large radii; because CRP5 has a gradual ramp up to a larger diffusion coefficient, its relative brightness at large radii is lower than the constant diffusion coefficient case of CRP4. This is also clearly shown in the radial intensity profile of fig.~\ref{fig:crp_radial_intensity}.

Our results for cosmic-ray protons indicate, as somewhat expected, that in order to support significant emission beyond the inner region, a comparatively small diffusion coefficient needs to be present in the outer regions of M31, as in CRP4 and CRP5. In either case, including when, as shown in fig.~\ref{fig:diffcoeff}, the diffusion coefficient is almost five times larger at the outskirts than around the inner M31 regions, the gamma-ray emission in the spherical halo and that in the inner region are quite accurately reproduced. Only a constant, suppressed diffusion coefficient would explain the outermost gamma-ray emission; in this case the emission in the inner region would also additionally be self-consistently explained (see the pink dashed line in fig.~\ref{fig:crp_radial_intensity}).

\section{Pulsar Emission}\label{sec:psr_gamma}

The pulsar population in M31 is modeled using the population synthesis code PsrPopPy \citep{psrpoppy}. 
In total we generate $10,000$ pulsars using default parameters for a Milky Way-type galaxy, appropriate in the present case. In particular, the radial distribution is based on the analysis of \cite{lorimer06}, while the vertical distribution assumes a two-sided exponential with scale height of $z_{scale}=0.33$ kpc. The pulsar spin period also is given in \cite{lorimer06} as a log-normal distribution with $\mu=2.7$, and standard deviation $\sigma=-0.34$. The total pulsar number was chosen as a reasonable estimate based on estimates from \cite{Lorimer2009}, where the theorized pulsar birth rate is estimated between 2.8 pulsars per century, which would give a population of approximately 3000 pulsars, and a larger possible number, which \cite{Lorimer2009} suggests could be up to a factor 5 larger. 

We sample the PSR ages homogeneously and linearly between the ages of $10^3$ yrs and $10^5$ yrs, to bracket the observationally-motivated age range of pulsars exhibiting a wind nebula (PWN) \cite{HESS2018}. 
As far as the size of the PWN, we implement the functional dependence between age and radius of the nebula as  in \cite{HESS2018, Profumo:2018fmz}. In figure \ref{fig:psr_locations} we show for illustrative purposes an image of the pulsar locations relative to the plane of the galaxy compared with a de-projected 24 $\mu$m image of M31 from \cite{Gordon_2006}. Notice that in the figure we use different scales for the $z$ axis and for the galactic plane. The right panel of fig.~\ref{fig:psr_locations} shows the size of the region of inefficient diffusion for a down-selection of 303 random pulsars, enhanced for visibility by a factor 20 in size.

\begin{figure*}
    \centering
    \includegraphics[width=0.85\linewidth]{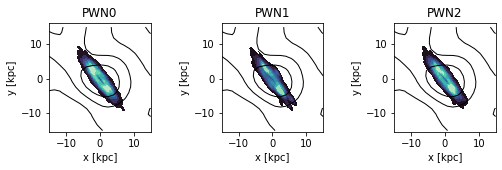}
    \caption{The morphology of the gamma-ray emission from unresolved gamma-ray pulsars as predicted in the population synthesis model we constructed, for the case where each pulsar features the same gamma-ray luminosity (PWN0), for  $L_\gamma\propto \dot E\propto 1/(\tau P^2)$ (PWN1), $L_\gamma\propto \sqrt{\dot E}\propto 1/(\sqrt{\tau} P)$ (PWN2).}
    \label{fig:psr_morphology}
\end{figure*}
\begin{figure}
    \centering
    \includegraphics[width=0.8\linewidth]{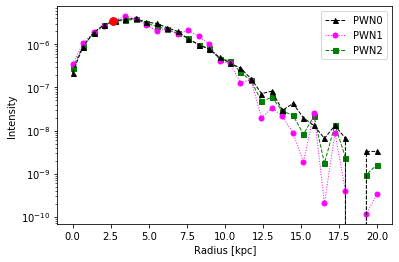}
    \caption{The radial intensity profile, normalized to the measured gamma-ray intensity from the inner regions of M31, for the gamma-ray emission from unresolved gamma-ray pulsars as predicted in the population synthesis model we constructed, for the case PWN0 of identical emission from each pulsar (black triangles), of $L_\gamma\propto \dot E\propto 1/(\tau P^2)$ (PWN1, purple circles), $L_\gamma\propto \sqrt{\dot E}\propto 1/(\sqrt{\tau} P)$ (PWN2, green squares).}
    \label{fig:psr_radial}
\end{figure}

We utilize the synthetic population constructed as described above to simulate the sources of high-energy CRE in the case of CRE6, and we adopt for diffusion scenario (iii), again from CRE6 and for CRP6$^\prime$ a suppressed diffusion coefficient 100 times smaller than in the rest of the diffusive cylindrical inner region. We also use the synthetic pulsar population to model a further possible (and plausible) source of gamma radiation.

As detailed in \cite{TheFermi-LAT:2013ssa}, the emission from gamma-ray pulsars is found to correlate with the pulsar's spin-down luminosity $\dot E$. Specifically, \cite{TheFermi-LAT:2013ssa} finds that, while several pulsars over a wide range of spin-down luminosities exhibit gamma-ray luminosities $L_\gamma\propto \dot E$, numerous other follow a phenomenological behavior where $L_\gamma\propto \sqrt{\dot E}$. Noting that the spin-down luminosity $\dot E\propto \dot P/P^3$, where $P$ is the pulsar period and $\dot P$ the period derivative, and that the characteristic pulsar age $\tau=P/(2\dot P)$, given our synthetic pulsar catalog, we simulated pulsar emission from the following two observationally motivated \cite{TheFermi-LAT:2013ssa} prescriptions:
\begin{itemize}
    \item [(1)] $L_\gamma\propto \dot E\propto 1/(\tau P^2)$;
    \item [(2)] $L_\gamma\propto \sqrt{\dot E}\propto 1/(\sqrt{\tau} P)$.
\end{itemize}
We also considered a model (0) where all pulsars produce the {\em same} gamma-ray emission. This latter case (0) can be considered a proxy for the emission from older, ``recycled'' millisecond pulsars, as considered for instance in \cite{eckner} and in \cite{fragione}.

We show in fig.~\ref{fig:psr_morphology} the expected gamma-ray emission from unresolved gamma-ray pulsars from the synthetic population we built as described above, for the three cases PWN1 where $L_\gamma\propto \dot E\propto 1/(\tau P^2)$, PWN2 where $L_\gamma\propto \sqrt{\dot E}\propto 1/(\sqrt{\tau} P)$ and in the case PWN0 where the same gamma-ray luminosity is associated with every pulsar in the catalog. We observe a slight increase to the extension of the emission from PWN1 to PWN2 to the same-luminosity case. Overall, however, the pulsar emission appears to be relevant only for the innermost few kpc, as also reproduced in the detailed radial intensity profile shown in fig.~\ref{fig:psr_radial}, where we normalize the emission to the inner galaxy data point, shown in red. 

We estimate that in order to account for 100\% of the inner galaxy emission, the typical gamma-ray luminosity of each pulsar in our catalog would need to be around $4\times 10^{40}$ ph sec$^{-1}$. Since the gamma-ray luminosity of both gamma-ray pulsars from the Fermi-LAT pulsar catalog \citep{TheFermi-LAT:2013ssa} and for millisecond pulsar \citep[see e.g.][]{Hooper:2015jlu} is $10^{33}-10^{37}$ erg/sec, and given that the observed emission peaks around 1 GeV, we find that the energetics of the gamma-ray emission is compatible with being primarily or in significant part fuelled by emission from gamma-ray pulsars. In summary, based on both morphological arguments and energetics, we find that emission from unresolved gamma-ray pulsars in M31 is likely to be a significant contributor in the inner regions of M31.

\section{Discussion and Conclusions}\label{sec:conclusions}
We studied the gamma-ray emission expected from the Andromeda galaxy (M31) due to high-energy cosmic-ray electrons and protons and by unresolved pulsar emission. The key motivation for the present study is the detection of gamma-ray emission in the spherical halo and far outer halo of M31 \citep{M31Halo}, and the possibility that such emission be in part or entirely associated with a cosmic-ray ``halo'' extending significantly beyond the disk and bulge of M31. 

We considered a broad ensemble of diffusion scenarios, including ones where diffusion is relatively efficient out to large radii, and ones where diffusion is significantly inhomogeneous. We found that cosmic-ray electrons up-scattering cosmic microwave background photons is likely responsible for a significant portion of the inner region gamma-ray emission and, possibly, of the spherical halo, especially if diffusion is highly inefficient near the sites of cosmic-ray electron acceleration. Cosmic-ray protons also definitely contribute to the inner-region emission, and possibly to the emission in the outer-most region, if the increase in the diffusion coefficient from the inner regions out to the virial radius is limited to within a factor 5-10. Finally, we studied the possible contribution of unresolved point-like sources associated with pulsars, and found that this should only contribute to the inner region, with limited impact on the spherical halo emission.

In this study we did not consider spectral information in investigating the nature of the spherical halo and outer halo emission from M31. The reason for this choice is that, as shown explicitly in \cite{McDaniel:2019niq}, in both the hadronic and the leptonic case the gamma-ray spectrum is largely dependent on the assumed cosmic-ray injection spectrum. The latter, in turn, depends on the acceleration sites and mechanism. As a result, while some information on the expected cosmic-ray spectrum is inferred from direct measurements at Earth, and from Galactic gamma rays, spectral information provides only a highly model-dependent input to the origin of the M31 gamma-ray emission.

In conclusion, our results suggest a possible direct evidence for an extended cosmic-ray halo around M31, and thus possibly around our own Galaxy, as first entertained for instance in \cite{Feldmann:2012rx} and \cite{Pshirkov:2015oqu}. The composition of the halo in terms of its leptonic and hadronic components is at present unclear, as is the detailed structure of the diffusion scenario, although our results give some boundaries to the nature of the latter. Given the possible new-physics interpretation of the gamma-ray emission from the outer regions of M31 \citep{M31HaloDM}, the cosmic-ray halo scenario should be carefully explored. Future observations both at gamma-ray and at other frequencies in conjunction with additional detailed cosmic-ray simulations and predictions at other wavelengths (including e.g. radio and X-ray, where detailed data exist) will help further elucidate the origin of the gamma-ray emission from M31. For instance, \cite{McDaniel:2019niq} showed that the inner galaxy M31 gamma-ray emission can only be explained exclusively by cosmic-ray electrons as long as the magnetic field in the inner regions is highly suppressed compared to expected values in the several micro-gauss range. Finally, given the similarities between M31 and the Milky Way, our results warrant establishing whether 
our own Galaxy possesses an extended cosmic-ray halo and, if so, how it would manifest observationally \citep[see e.g.][]{Tibaldo:2015ooa}.

\section*{Acknowledgements}
SP is partly supported  by the U.S. Department of Energy grant numbers DE-SC0010107 and A00-1465-001. We gratefully acknowledge early contributions by Henri Geneste, Lili Manzo, Lilianne Callahan, and helpful conversations with Tesla Jeltema.

\section*{Data Availability}
The data underlying this article will be shared on reasonable request to the corresponding author.

\bibliographystyle{mnras}
\bibliography{main}
\label{lastpage}
\end{document}